\documentclass[
 reprint,
 amsmath,amssymb,superscriptaddress,
 aps,pra
]{revtex4-1}

\usepackage{graphicx}% Include figure files
\usepackage{dcolumn}% Align table columns on decimal point
\usepackage{bm}% bold math
\usepackage{hyperref}% add hypertext capabilities
\hypersetup{colorlinks=true, citecolor=blue}
\usepackage[caption=false]{subfig}
\usepackage{physics}
\usepackage{bbm}
\usepackage{amsmath}
\newcommand{\crea}[1]{\hat{#1}^{\dagger}}

\begin{document}

%\preprint{APS/123-QED}

\title{Violating the Leggett-Garg inequalities with classical light}% Force line breaks with \\
\author{Hadrien Chevalier}\thanks{hadrien.chevalier17@imperial.ac.uk}
\author{A. J. Paige}\thanks{a.paige16@imperial.ac.uk}
\author{Hyukjoon Kwon}\thanks{h.kwon@imperial.ac.uk}
\affiliation{QOLS, Blackett Laboratory, Imperial College London, SW7 2AZ, United Kingdom}
\author{M. S. Kim}\thanks{m.kim@imperial.ac.uk}
\affiliation{QOLS, Blackett Laboratory, Imperial College London, SW7 2AZ, United Kingdom}
\affiliation{Korea Institute of Advanced Study, Seoul, 02455, Korea}
\date{\today}% It is always \today, today,

\begin{abstract}
In an endeavour to better define the distinction between classical macroscopic and quantum microscopic regimes, the Leggett-Garg inequalities were established as a test of macroscopic-realistic theories, which are commonly thought to be a suitable class of descriptions for classical dynamics. The relationship between their violation and non-classicality is however not obvious. We show that classical states of light, which in the quantum optical sense are any convex sums of coherent states, may not satisfy the Leggett-Garg inequalities. After introducing a simple Mach-Zehnder setup and showing how to obtain a violation with a single photon using negative measurements, we focus on classical states of light, in particular those of low average photon number. We demonstrate how one can still perform negative measurements with an appropriate assignment of variables, and show that the inequalities are violable with coherent states. Finally, we abandon initial phase reference and demonstrate that the violation is still possible, in particular with thermal states of light, and we investigate the effect of intermediate dephasing.
%\begin{description}
%use the optional argument of the %\verb+\item+ command to give the category of each item. 
%\end{description}
\end{abstract}

\pacs{Valid PACS appear here}% PACS, the Physics and Astronomy
\maketitle

%\tableofcontents

\section{\label{sec:level1}Introduction}

One of the hallmarks of quantum theory is the coherent superposition of states. States which have coherence~\cite{winter2016operational, theurer2017resource} are an important resource in applied physics, notably for the development of quantum information and its applications~\cite{chuang}, such as metrology~\cite{giovannetti2004quantum, giovannetti2006quantum, giovannetti2011advances} or computation~\cite{duan1997preserving, steane1998quantum, jeong2002efficient, hillery2016coherence, ma2016converting}. However, coherent superpositions have long been at the core of fundamental issues, famously illustrated by Schr\"{o}dinger's cat gedankenexperiment~\cite{schrodinger1935gegenwartige}. Making sense of the disconnect between quantum microscopic and classical macroscopic regimes has given rise to different models for and interpretations of open-system quantum dynamics~\cite{schlosshauer2005decoherence, holland1995quantum, ballentine1970statistical, dewitt2015many, cramer1986transactional}.

Similar to Bell's inequalities, based on correlations between spatially separated systems, which are a necessary condition for local realism~\cite{bell1964einstein}, the Leggett-Garg inequalities (LGIs) test the validity of classical descriptions through the correlations between successive measurements in time of a single system~\cite{leggettoriginal}. The LGIs are a consequence of macroscopic realism, or macrorealism (MR), which is defined as the conjunction of three assumptions~\cite{emaryreview}: macroscopic realism \textit{per se} {(MRps)}, under which a system which has access to a number of distinct macroscopic states is always in exactly one of those states; non-invasive measurability {(NIM)} under which a system's state can be measured by impinging an arbitrarily faint perturbation on that state; and induction (I) which demands that information be forward propagating in time. Induction is usually taken for granted.

Since the LGI was first proposed~\cite{leggettoriginal}, several subtleties regarding what is meant by MR and the precise significance of an LGI violation have been clarified~\cite{moreira2015modeling,halliwell2019necessary}. Contrary to the no-signaling in time conditions which are equivalent to MR, the LGI is a necessary condition for MR \cite{halliwell2017comparing, clemente2016no}. It was furthermore pointed out that there exist different varieties of MRps, and the only flavour of MRps that can be ruled out by an LGI violation is termed ``operational eigenstate mixture macrorealism''~\cite{maroney2014quantum}. Hence this is the meaning of MRps adopted in our work. LGI violations do not rule out the existence of a hidden variable model explaining the violation~\cite{katiyar2017experimental}. However, they have long been considered to be relevant witnesses of non-classicality~\cite{lambert2011macrorealism,li2012witnessing}, and finding such violations is still an active area of interest~\cite{huffman2017violation, naikoo2019violation, thenabadu2019leggett, zhang2020influence, bose2018nonclassicality, halliwell2020leggett}. To no surprise, the LGIs have been experimentally violated with microscopic systems such as superconducting qubits and atomic quantum walks~\cite{leggett2008,robens2015ideal}.

There exists another well established notion of classicality of a state, in quantum optics. A state of light $\rho$ can be represented by a distribution in the complex plane as $\rho = \int P(\alpha)\dyad{\alpha} d^2\alpha$ where $\ket{\alpha}$ are coherent states~\cite{glauber1963coherent, sudarshan1963equivalence}. The state is said to be classical whenever $P$ is a probability density function on phase space~\cite{titulaer1965correlation}. This criterion is justified by the fact that a coherent state is considered a classical pure state~\cite{mandel1986non}, in the sense that it minimizes uncertainty relations and is robust against decoherence~\cite{zurek1993coherent}. By contrast, a superposition of coherent states $\ket{\alpha}$ and $\ket{\beta}$, is non-classical, and when the displacement parameters $\alpha,\beta$ differ considerably, this is referred to as a Schr\"{o}dinger cat state~\cite{ourjoumtsev2007generation}. Such a non-classical state, whose $P$ representation is not a probability density function, is a valuable resource for quantum information tasks~\cite{yadin2018operational,kwon2019nonclassicality}, as indicators of quantum behaviour. Yet, it turns out that classical states can very well exhibit quantum properties, especially when they have a significant vacuum component, as we shall show.

In this work we investigate whether states of light that have positive $P$-functions, which we refer to as classical states, such as coherent and thermal states, can simulate quantum behaviour, specifically the violation of an LGI. Violations with light have already been established~\cite{goggin2011violation, xu2011experimental, dressel2011experimental, suzuki2012violation, wang2018violations,wang2020experimental}, although with manifestly non-classical states such as single photons. Violations using the polarization degree of freedom of a laser field were more recently shown to be possible~\cite{zhang2018experimental}. However this violation is a particular implementation of a qubit to violate the LGI, whereas our proposal uses measurements on the coherent state itself to achieve the violation. More importantly, the previously established violations have not determined to what degree phase reference, which plays a central role in the decoherence model explaining the quantum-classical transition~\cite{zeh1970interpretation}, is necessary to have LGI violations. We will demonstrate a violation of the LGI with a particularly simple setup in which light is classical at each stage and the measurement itself is not weak~\cite{aharonov1988result}. 

Our work is organized as follows. We begin by showing a known derivation of the LGI and we introduce the Mach-Zehnder setups to establish an LGI violation with a single photon Fock state in Sec.~\ref{sec:singlephoton}. We suggest in Sec.~\ref{sec:coherent} a new observable assignment for non-dichotomic variables to permit negative measurements. This allows one to show how an LGI violation can be obtained with coherent states of light. Finally, we investigate the effects of losing phase reference. We consider dephased input states in Sec.~\ref{sec:dephased} and demonstrate that LGI violations are still possible, namely with thermal states of light. We also show in Sec.~\ref{sec:decoh} that complete intermediate dephasing prevents any LGI violation.

\begin{figure*}
\centering
\subfloat[Detector on the right intermediate mode.\label{sfig:MZIR}]{%
  \includegraphics[height=8cm,width=.32\linewidth]{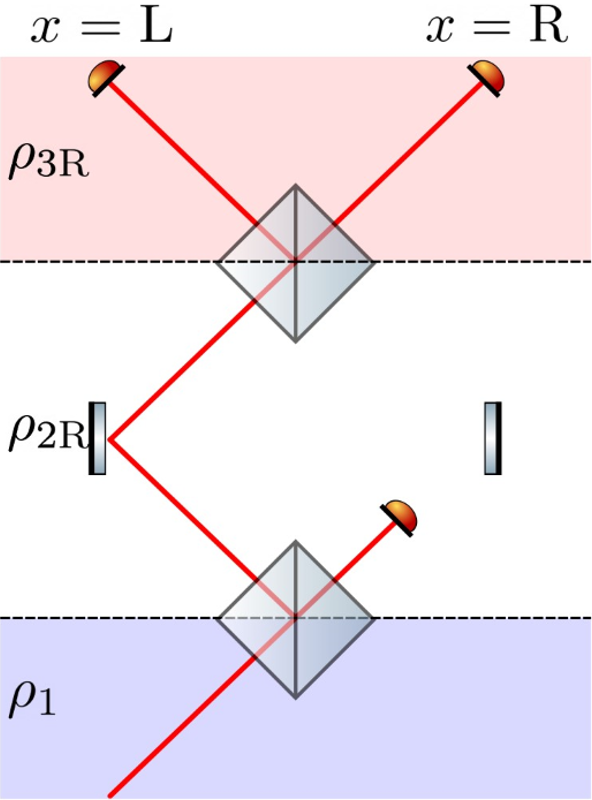}}%
\subfloat[No intermediate detection.\label{sfig:MZI}]{%
  \includegraphics[height=8cm,width=.32\linewidth]{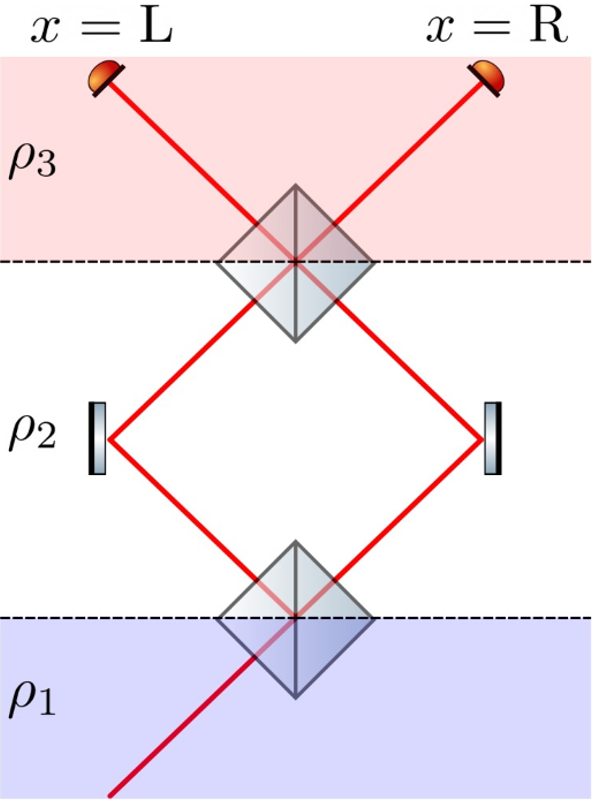}}%
\subfloat[Detector on the left intermediate mode.\label{sfig:MZIL}]{%
  \includegraphics[height=8cm,width=.32\linewidth]{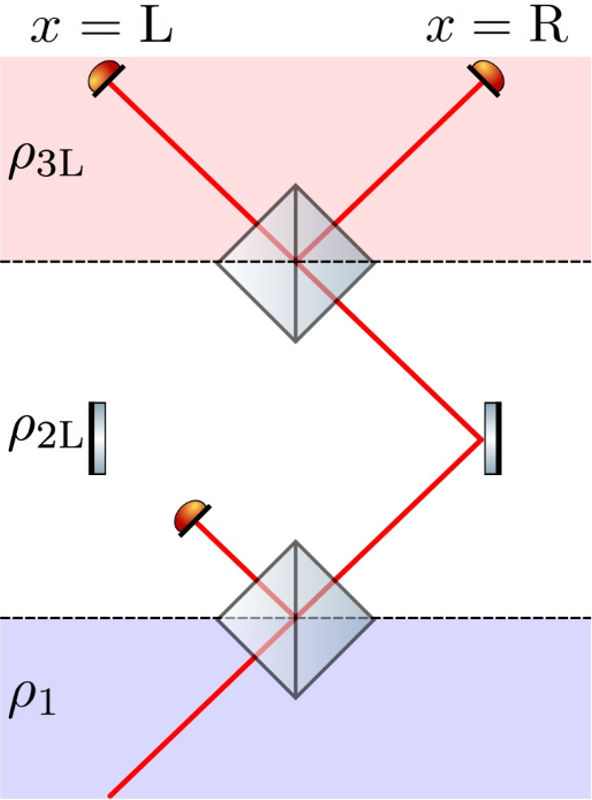}}%

\caption{Three setups for the two step quantum random walk. The states in each space, input, intermediate, and output, are labelled according to the intermediate detector's position. Photons are detected at the output of the Mach-Zehnder interferometer at two distinct positions $x=L$ and $x=R$.}
\label{}
\end{figure*}

\section{\label{sec:singlephoton}LGI violation with a single photon}
Before introducing the setup and illustrating an LGI violation, let us briefly show how to establish the LGIs, and introduce useful notations as well as the notion of negative measurement. Following review~\cite{emaryreview} we present a brief derivation of the LGIs.

Using the ontic models framework~\cite{spekkens2005contextuality}, let us suppose that the system under study is prepared in the ontic state $\sigma$ with a probability density $\pi(\sigma)$. A measurement that is performed at instant $t_i$ results in the outcome function $\mu_i(Q_i|\sigma)$ which gives the probability of obtaining the value $Q_i$ given the ontic state $\sigma$. Induction guarantees that preparing a measurement setup does not influence the initial ontic state distribution $\pi(\sigma)$. Then $\mathbb{P}(Q_i) = \int d\sigma \mu_i(Q_i|\sigma)$. Assuming NIM, the joint probability distribution for the two measurements reads $\mathbb{P}_{ij}(Q_i,Q_j) = \int d\sigma \mu_j(Q_j|\sigma)\mu_i(Q_i|\sigma)\pi(\sigma)$. Let us now restrict to $(Q_i,Q_j)\in S_i\times S_j \subset [-1,1]^2$. $S_i$ and $S_j$ are the sets of values that $Q_i$ and $Q_j$ can respectively take, and those values are real and absolutely less than or equal to unit. Then the correlation coefficient reads $C_{ij} = \ev{Q_iQ_j} = \sum_{(Q_i,Q_j)\in S_i\times S_j} Q_iQ_j\mathbb{P}_{ij}(Q_i,Q_j)$. Inserting the joint probability expression into the last equation gives $C_{ij} = \int d\sigma \ev{Q_i}_{\sigma}\ev{Q_j}_{\sigma}\pi(\sigma)$, where the expectation value is explicitly $\ev{Q}_{\sigma} = \sum_{Q\in S} Q\mathbb{P}(Q)$. Thus $K = C_{12} + C_{23} - C_{13} = \int d\sigma [\ev{Q_1}_{\sigma}\ev{Q_2}_{\sigma} + \ev{Q_2}_{\sigma}\ev{Q_3}_{\sigma} - \ev{Q_1}_{\sigma}\ev{Q_3}_{\sigma} ] \pi(\sigma)$.
Therefore, having $S_1\times S_2\times S_3\subset [-1,1]^3$ yields
\begin{equation}
    K = C_{12} + C_{23} - C_{13} \leq 1,
    \label{eq:LGIgeneral}
\end{equation}
regardless of the cardinality of the sets $S_i$. In particular, it will be useful in Sec.~\ref{sec:coherent} to note that if $(S_1, S_2, S_3) = (\{+1\}, \{0,+1\}, \{-1, 0, +1\})$ then the LGI~\eqref{eq:LGIgeneral} still holds.

Let us now show a simple violation in a Mach-Zehnder interferometer. Mach-Zehnder setups have been considered to test LGIs with dichotomic variables~\cite{kofler2013condition}, and featured for detailed proposals of LGI violations with electrons~\cite{emary2012leggett, emary2012leggett2}. We present the three Mach-Zehnder setups, along with the notations and observable definition, and give an example of LGI violation with macroscopic observables and negative measurement in the case of a single photon input state.

We consider three setups as illustrated in figures \ref{sfig:MZIR}, \ref{sfig:MZI} and \ref{sfig:MZIL}. All in all, the ideal setup consists of two perfect mirrors, two or three photon detectors and two identical 50:50 lossless beam splitters. For our purpose the detectors need not be photon counters, but rather simply detect the presence or absence of photons in the mode. The general beam splitter operator is defined as $\Hat{B} = e^{\frac{\theta}{2}(\crea{a}_{\text{L}}\hat{a}_{\text{R}} - \hat{a}_{\text{L}}\crea{a}_{\text{R}})}$ where $\crea{a}_{\text{L}}$ and $\crea{a}_{\text{R}}$ are bosonic creation operators in the left-hand and right-hand field modes, and we shall fix $\theta = \pi/2$ throughout this paper. Using Hadamard's lemma and bosonic commutation relations, the 50:50 beam splitter acts upon photonic modes according to the following rules:
\begin{equation}
    \left\{\begin{array}{ccc}
        \Hat{B}\crea{a}_{\text{R}}\crea{B} & = & \frac{1}{\sqrt{2}}(\crea{a}_{\text{L}} + \crea{a}_{\text{R}}) \\
         \Hat{B}\crea{a}_{\text{L}}\crea{B} & = & \frac{1}{\sqrt{2}}(\crea{a}_{\text{L}} - \crea{a}_{\text{R}})  
    \end{array}\right..
    \label{eq:BStransformations}
\end{equation}
There are two setups for the intermediate measurement, and this is a requirement for ideal negative measurements, also known as interaction-free measurements~\cite{elitzur1993quantum, kwiat1995interaction}. Such measurements are important in order to have a meaningful LGI violation, as direct measurements disturb the state and immediately invalidate the NIM hypothesis. The idea of a negative measurement, in the single photon case, is to say that by not observing a photon in one of the two detectors, one can conclude its presence in the other mode without having destroyed it. If an intermediate detector clicks, the trial is discarded, but this case is accounted for when the detector is in the other mode. Of course, even negative measurements do disturb the quantum state, however from a realist's point of view, it is but an update of an agent's knowledge of the state of the system. 

Let us note that this measurement method is straightforward only when the beam splitters are lossless and the detectors are ideal (no dark current, and unit quantum efficiency), which we assume in this work. We briefly discuss in the next section why this assumption does not prevent our proposal from being viable.

This being said, let us consider a single photon arriving on the first beam splitter from the left, so $\ket{\psi_1} = \ket{10}$ is the input state. With no intermediate detection, the intermediate state between the two beam splitters is the Bell pair $\ket{\psi_2} = \frac{1}{\sqrt{2}}(\ket{10} - \ket{01})$, and the output state is $\ket{\psi_3} = -\ket{01}$. If the intermediate detector is placed on the right hand intermediate mode, then the only state that one can measure negatively is $\ket{\psi_2}_{\text{R}} = \ket{10}$. Similarly, we will denote $\ket{\psi_2}_{\text{L}} = -\ket{01}$ the only negatively measurable state when the detector is placed on the left intermediate mode. 

Those are negatively measured states in the sense that one can deduce their form from the absence of a click on the intermediate detector, which is a pivotal point to make in order to even consider NIM. Hence all trials where the intermediate detector clicks have to be discarded. The output states after negative intermediate measurement are then $\ket{\psi_3}_{\text{R}} = \frac{1}{\sqrt{2}}(\ket{10} - \ket{01})$ and $\ket{\psi_3}_{\text{L}} = \frac{1}{\sqrt{2}}(\ket{10} + \ket{01})$, where the index is a label for the position of the intermediate detector, and is absent if no intermediate measurement is performed.

It should be stressed that discarding trials where the intermediate measurement was not negative (i.e. a detector clicked) does not affect the resulting statistics only if the discarded cases can be picked up in the statistics in the symmetric setup, as illustrated with atomic quantum random walks in~\cite{robens2015ideal}. %The fact that there are many intermediate detection setups to recover the discarded trials is crucial to the usage of negative measurements in the aim of having NIM.

We now define the $Q_i$ values to obtain a violation as follows. We set $Q_1 = +1$ when there is a photon in the left input mode and none in the right input mode. This corresponds to the preparation. We trivially set $Q_2 = +1$ whenever the photon finds itself in either of the intermediate modes. Finally, we set $Q_3 = +1$ when the left output detector clicks, and $Q_3 = -1$ if the right output detector clicks. 

Then it is straightforward to establish $\ev{Q_3} = -1$ and $\ev{Q_3}_{\text{R,L}} = 0$. Trivially $C_{12} = +1$, and $C_{13} = \ev{Q_3} = -1$. Finally, as $Q_2 = +1$ always holds and trials where the intermediate detector clicks are discarded (i.e. half of the trials for each intermediate detector position), one has $C_{23} = \frac{1}{2}(\ev{Q_3}_{\text{R}} + \ev{Q_3}_{\text{L}})$. All in all $K = +2$ which violates~\eqref{eq:LGIgeneral}.

\section{LGI violation with a coherent state}\label{sec:coherent}

We now consider a coherent state impinging on the first beam splitter from the left $\ket{\psi_1} = \ket{\alpha}_{\text{L}}\ket{0}_{\text{R}} = \hat{D}_{\text{L}}(\alpha)\otimes \mathbbm{1}_{\text{R}}\ket{00}$, where $\Hat{D}(\alpha) = e^{\alpha\crea{a} - \alpha^*\Hat{a}}$ is the displacement operator. At the output of the first beam splitter, the state of light is $\ket{\psi_2} =\ket{\alpha/\sqrt{2}}\otimes \ket{-\alpha/\sqrt{2}}$. Hence, the state at the output of the interferometer with no intermediate measurement is given by $\ket{\psi_3} = \ket{0}\otimes\ket{-\alpha}$. In setups with intermediate measurement where any detected flux results in a discarded experiment, the negatively obtained states at the output are $\ket{\psi_3}_{\text{R,L}} = \ket{\pm\alpha/\sqrt{2}}\otimes\ket{\pm\alpha/\sqrt{2}}$.

Given the different output states, it may seem at first sight that the LGI violation will immediately follow from what was already shown for the single photon. However, the initial assignment for the observables $Q_i$ that lead to the single photon LGI violation, despite being a good starting point, is problematic. The issue is twofold: the observables as defined previously are no longer well determined, and their values can no longer be negatively measured with a state-selective discarding. Let us explicitly make those points and present a solution.

First, keeping $Q_3 = +1$ when photons impinge on the left detector at the output, and none to the right, and $Q_3 = -1$ when photons impinge to the right and none to the left, would result in an observable that does not have distinct states. Since there is not just a single photon, both modes could carry photons at the same time, in which case $Q_3$ would have two simultaneous values.

The second point is more troublesome. While with a single photon, trials that are discarded are picked up in the statistics using the symmetric setup, this no longer holds with multiple photons. Indeed, when a flux is detected at $t_2$ then two cases arise: either there are no photons in the other mode, or there are.

In the first case, the discarded trial is accounted for in the symmetric setup. In the second case, however, the trials in which there were photons in both output modes of the first beam splitter are simply lost. This poses an issue with non-invasiveness as artificially selecting only cases where all the flux is in one mode would completely alter the $C_{23}$ correlator.

 We propose a way to solve this issue by choosing the set of values for $Q_2$ to be $\{0, +1\}$. In particular, we include $0$ specifically as a possible value and will make use of its annihilating property. The assignments are summed up in Tab.~\ref{fig:assignment} and the corresponding quantum operators for observables $Q_2$ and $Q_3$ are shown in Eqs.~\eqref{eq:q2def} and~\eqref{eq:q3def}. Such an assignment was obtained as follows.
\begin{table}[h!]
    \centering
    \begin{tabular}{|c|c||c|c|c|} \hline
    L & R & $Q_1$ & $Q_2$ & $Q_3$ \\ \hline
    Vacuum & Vacuum & $+1$ & $0$ & $-1$ \\ \hline
    Vacuum & Photons & N.A. & $+1$ & $-1$ \\ \hline
    Photons & Vacuum & $+1$ & $+1$ & $+1$ \\ \hline
    Photons & Photons & N.A. & $0$ & $0$ \\ \hline
\end{tabular}
    \caption{Assignment of values for the observables, with respect to mode states. $L$ and $R$ designate modes that are respectively on the left hand side and on the right hand side. For $Q_1$ those are the input modes, for $Q_2$ those are the intermediate modes and for $Q_3$, the output modes. As the preparation of the experiment sets the right input mode in the vacuum state, no values need to be assigned in other cases for $Q_1$, though any arbitrary value would be valid.}
    \label{fig:assignment}
\end{table}

We set $Q_1 = +1$ when there are no photons in the right input mode. Other cases concerning the first beam splitter's input states never occur as this is the way the experiment is prepared. The preparation process results in $Q_1 = +1$ constantly.

We furthermore set $Q_3 = -1$ when the left detector does not click, $Q_3 = +1$ when the right detector does not click and the left detector clicks, and $Q_3 = 0$ when both detectors click. Note that when no detectors click at all, $Q_3 = -1$.

Finally we must be careful with the assignment of $Q_2$.  In an attempt to salvage non-invasive measurability, we will define $Q_2 = +1$ when all photons are in the same intermediate mode. Perhaps the most important choice, and what saves the negative measurement method, is the assignment $Q_2 = 0$ if there are photons in both intermediate modes or if there are no photons at all. This way, $Q_2 = +1$ will be realized as long as exactly one of the two intermediate modes is in the vacuum state, and $Q_2 = 0$ otherwise.

Setting $Q_2 = 0$ in the case where there are photons in both modes may seem to make it so that the discarded trials, which are permanently lost, in fact would not have had any impact on the $C_{23}$ correlation coefficient as $C_{23} = \ev{Q_2Q_3} = \sum Q_2Q_3\mathbb{P}_{23}(Q_2,Q_3)$. So, be they discarded or not, instances in which photons are in both intermediate modes would not contribute to $C_{23}$.

However, this reasoning is too hasty, as discarding cases when $Q_2 = 0$, while having no effect on the number of nonzero terms in the sum defining $C_{23}$, does nonetheless change the joint probability distribution $\mathbb{P}(Q_2,Q_3)$. %Indeed, discarding trials where $Q_2 = 0$ sets $\mathbb{P}(0,Q_3) = 0$ and so increases the values of $\mathbb{P}(Q_2\neq 0, Q_3)$.

Nevertheless, setting $Q_2 = 0$ when both intermediate modes contain photons does make it possible to save non-invasive measurability, but in fact without discarding any trials that cannot be negatively distinguished. To show this, let us observe the four following cases that make up all possible situations :\\
\begin{enumerate}
    \item If the detector at $t_2$ does not click, then
        \begin{enumerate}
            \item Either no photons are detected at $t_3$ which means there were no photons at all so $Q_2 = 0$.
            \item Or photons are detected at $t_3$, so $Q_2 = +1$ because all photons are in the other mode, then $Q_2 = +1$ is known via a negative measurement.
        \end{enumerate}
    \item If the detector at $t_2$ does click then
        \begin{enumerate}
            \item Either no detectors click at $t_3$ in which case $Q_2 = +1$ but we can discard the trial, and this situation is taken into account in the other setup where the intermediate detector is positioned on the other mode.
            \item Or detectors do click at $t_3$ in which case $Q_2 = 0$ because there were photons in both modes. Then by having chosen $Q_2 = 0$ we do not need to discard this trial, as regardless of the measurement outcome at time $t_3$ one will have $Q_2Q_3 = 0$ in any case. This means that whether the $t_2$ measurement was invasive or not does not matter at all. Whether the $Q_3$ value that is obtained was a possessed value or a measured value plays no role either. What matters is that at least one detector clicks at $t_3$, but the measurement outcome value is of no importance.
        \end{enumerate}
\end{enumerate} \mbox{}\\
\indent In this manner, direct invasive measurements at $t_2$ are either discarded but not permanently lost, or a rigorously non-invasive $t_2$ measurement would contribute to $C_{23}$ in the exact same way as the possibly invasive real $t_2$ measurement. Another way of phrasing what we have done is that by setting $Q_2 = 0$ when there are photons in both intermediate modes, the only measurements that contribute to $C_{23}$ are either negative measurements or effectively non-invasive measurements. All in all, the whole argument to salvage non-invasive measurability hinges on the use of the value $0$ which is absorbing (or annihilating) for the multiplication. For clarity we explicitly show of all the cases in the measurement of $C_{23}$ in Tab.~\ref{fig:cases}.

\begin{table}[h!]
    \centering
    \begin{tabular}{|c|c|c||c|c|c|} \hline
    $t_2$ & $t_3\text{L}$ & $t_3\text{R}$ & $Q_2$ & $Q_3$ & Case and $t_2$ measurement type \\ \hline
    $0$ & $0$ & $0$ &  $0$ & $-1$ & 1.(a) negative\\ \hline
    $0$ & $0$ & $1$ & $+1$ & $-1$ & 1.(b) negative \\ \hline
    $0$ & $1$ & $0$ & $+1$ & $+1$ & 1.(b) negative \\ \hline
    $0$ & $1$ & $1$ & $+1$ & $0$  & 1.(b) negative \\ \hline
    $1$ & $0$ & $0$ & $+1$ & $-1$ & 2.(a) direct and invasive (discarded) \\ \hline
    $1$ & $0$ & $1$ & $0$ & $-1$ & 2.(b) direct and effectively non-invasive\\ \hline
    $1$ & $1$ & $0$ & $0$ & $+1$ & 2.(b) direct and effectively non-invasive\\ \hline
    $1$ & $1$ & $1$ & $0$ & $0$ & 2.(b) direct and effectively non-invasive\\ \hline
\end{tabular}
    \caption{All possible measurement outcomes in the evaluation of the $C_{23}$ correlator. Clicks are symbolized by $1$ and absence of clicks by $0$. The only discarded trials are the ones in which $Q_2 = +1$ is directly measured via the click of the $t_2$ detector but the absence of clicks at $t_3$. Those trials are not lost as they are counted in the symmetric setup, where the $t_2$ detector is placed in the other mode. All other direct measurements at $t_2$ (when the $t_2$ detector clicks) are effectively non-invasive and so are not discarded, as the $Q_2Q_3$ outcome no longer depends on the value measured at $Q_3$, as long as at least one of the $t_3$ detectors click.}
    \label{fig:cases}
\end{table}

One point worth discussing is the exposure to a fair sampling loophole. We have assumed here that all detectors are ideal, however our assignment does not absolutely require unit quantum efficiencies and noiseless dynamics. To show that the experiment can work in principle with imperfect detectors, consider an overall error rate $\eta$ and let us suppose all errors give the worst outcome (i.e. skews the average $K$ value the most towards an LGI violation) in which $Q_1Q_2 + Q_2Q_3 - Q_1Q_3 = 3$. Then under macroscopic realism the highest attainable value for $K$ is $(1 + 2\eta)$. If each of the three detectors used to establish the $Q$ values has a generic error rate (in telling apart the vacuum from a non-vacuum state) $\varepsilon$, then the overall error rate will be $\eta = 1 - (1 - \varepsilon)^3$. Taking $\varepsilon = 5\%$ yields $\eta = 0.15$ so that the $K$ function threshold for an LGI violation may be shifted to $1.3$. As we shall see, this new threshold can be exceeded with a coherent state input, as well as with a thermal field input.

Let us now show that the violation is indeed achieved. To this end, recall that the probability of detecting $n$ photons in a coherent state $\ket{\alpha}$ is Poissonian
\begin{equation}
    p_n(\alpha) = e^{-|\alpha|^2}\frac{|\alpha|^{2n}}{n!}
    \label{eq:poissonianprob}.
\end{equation}

\begin{figure}[h!]
    \centering
    \includegraphics[scale=0.55]{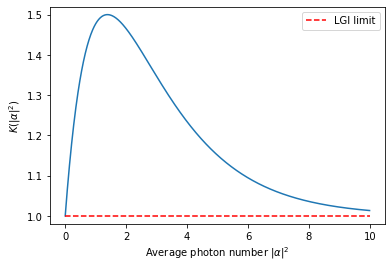}
    \caption{LG correlation function with respect to the average photon number. Plot obtained for a coherent state input in the Mach-Zehnder interferometer with appropriate observable value assignment.}
    \label{fig:Kfunction}
\end{figure}

First $C_{12} = \ev{Q_2}$ can be expressed by introducing the photon numbers $n_{\text{R}}$ and $n_{\text{L}}$ respectively in the right and left intermediate modes. We may rewrite the event $\{Q_2 = 0\}$ as $\big(\{n_{\text{L}} > 0\}\wedge\{n_{\text{R}} > 0\}\big)\vee\big(\{n_{\text{L}} = 0\}\wedge\{n_{\text{R}} = 0\}\big)$, and note that photon numbers in the two output modes are independent of one another. Hence using Eq.~\eqref{eq:poissonianprob} we obtain $C_{12} = \mathbb{P}(Q_2 = +1) = 4e^{-\frac{3|\alpha|^2}{4}}\sinh(|\alpha|^2/4)$. Next, $C_{13} = \ev{Q_3} = -1$ is straightforward as the output state when there is no intermediate measurement is  $\ket{0}\otimes\ket{-\alpha}$, hence all the flux arrives at the right hand output detector. Finally $C_{23} \propto \ev{Q_3}_{\text{L}} + \ev{Q_3}_{\text{R}} = 0$, because the beam splitters are 50:50. Since cases where possible interference may occur, and the vacuum, are assigned $Q_2 = 0$, the situations that contribute to $C_{23}$ are those where photons impinge on the second beam splitter from only one side. Then the detection probabilities are equal in both output modes, and since $Q_3$ takes opposite values in those cases, the average value is null. This results in the following LG correlation
\begin{equation}\label{eq:Kcoherent}
    K(\alpha) = 1 + 4e^{-\frac{3|\alpha|^2}{4}}\sinh(\frac{|\alpha|^2}{4}),
\end{equation}
for which a plot is shown in Fig.~\ref{fig:Kfunction}.

We observe a violation of the LGI with a maximum violation when the average photon number $|\alpha|^2$ is just over $1$. More explicitly, the maximum is reached at $|\alpha|^2 = 2\ln(2)$, with a value of $K(\alpha_{\text{max}}) = 1.5$. Let us note that this maximum is reached when the intermediate modes are equally balanced superpositions of the vacuum state and all other Fock states $2^{-1/2}(\ket{0} + \ket{n>0})$.

We also note that the $K$ function decays for high laser intensities. This is due to our observable value assignment choice. Indeed, as the laser field becomes more intense, trials in which all photons end up in the same mode become less likely, so we should expect $C_{12}$ to drop to $0$. 
This shows that even if classical light can in principle violate an LGI, the violation will become extremely unlikely to occur experimentally at high laser intensities.
%This violation suggests that classical wave mechanics may simulate LGI violations. %However, as quanta of energy can be detected and even counted, a realist might argue that the coherent state in its quantum description is not an accurate representation of what really is at the output of a laser. To close this eventual loophole we have taken it a step further, by letting go of any phase reference in the input state. In other words, we perform the calculations with realist-friendly states~\cite{maroney2014quantum}, in the sense that they are statistical mixtures of photon numbers.

\section{\label{sec:dephased}LGI violation with dephased input states}

Using the same setup and observable assignment, we let go of any sort of phase reference in the input coherent state. That is to say, we now consider the input state $\rho_1(\alpha) = \sum_{n=0}^{+\infty} p_n(\alpha)\dyad{n,0}{n,0}$, where $p_n(\alpha)$ is given by Eq.~\eqref{eq:poissonianprob}. The output state is formally given by
$\rho_3 = \hat{B}\hat{B}\rho_{1}\crea{B}\crea{B}$.
By linearity of all performed operations, we may as well simplify calculations by considering Fock input states
$\rho_1(n) = \dyad{n,0}{n,0}$. The state after the first beam splitter reads $\rho_2(n) = \frac{1}{n!}\Hat{B}\hat{a}^{\dagger\, n}_{\text{L}}\hat{B}^{\dagger}\dyad{0,0}{0,0}\Hat{B}\Hat{a}_{\text{L}}^n\hat{B}^{\dagger}$.
The beam splitter transformations~\eqref{eq:BStransformations} yield
\begin{equation}\rho_2(n) = \frac{1}{2^n}\sum_{k,\ell = 0}^n\sqrt{\binom{n}{k}\binom{n}{\ell}}(-1)^{k+\ell}\dyad{k, n-k}{\ell, n-\ell}.
\end{equation}
Let us begin by computing $C_{12} = \ev{Q_2}$ where
\begin{equation}\label{eq:q2def}
    \Hat{Q}_2 = \sum\limits_{n=1}^{+\infty} \dyad{0,n}{0,n} + \dyad{n,0}{n,0}.
\end{equation}
Since $Q_2 \in \{0, +1\}$ the expectation value is simply $\ev{Q_2} = \mathbb{P}(Q_2 = +1)$. Because we assume 50:50 beam splitters, we can write $\mathbb{P}(Q_2 = +1) = \sum_{n=1}^{+\infty} \frac{1}{2^{n-1}}p_n(\alpha) = 2e^{-|\alpha|^2}\left( e^{|\alpha|^2/2} -1\right)$.
This result, which can also be arrived at directly as shown in Appendix~\ref{sec:c12derivation} (with parameter $\gamma = 1$ as no decoherence is considered at this stage), is identical to the previously established expression for $C_{12}$.

As previously argued $C_{23} = 0$, by virtue of the following inspection. If photons are in both modes or there are no photons, then $Q_2 = 0$, so those cases, regardless of the obtained value of $Q_3$, do not contribute to $C_{23}$. If photons are all in the same mode then $Q_2 = +1$ and because the beam splitter is 50:50 the assignment of $Q_3$ values results in an overall average value of $0$.

Finally, to compute $C_{13}$, what we seek is $\ev{Q_3} = \Tr(\rho_3\Hat{Q}_3)$ where
\begin{equation}\label{eq:q3def}
    \Hat{Q}_3 = \left(\sum\limits_{n=1}^{+\infty} \dyad{n,0}{n,0} - \dyad{0,n}{0,n}\right) - \dyad{0,0}{0,0},
\end{equation}
so we may project out all components of the density operator for which the product with $\Hat{Q}_3$ would give an off-diagonal element. The projected density matrices will be written with a tilde $\tilde{\rho}$. Applying the beam splitter transformations \eqref{eq:BStransformations}, one finds the projected output state $\tilde{\rho}_3(n) = \dyad{0,n}{0,n}$, so that $\tilde{\rho}_3(\alpha) = \sum_{n=0}^{+\infty} p_n(\alpha)\dyad{0,n}{0,n}$. A detailed proof of the expression for $\tilde{\rho}_3(n)$ can be found in Appendix~\ref{sec:proofs}. From this, one finds $C_{13} = \ev{Q_3} = \Tr(\hat{Q}_3\tilde{\rho}_3(\alpha)) = -1$, which is identical to the previously established $C_{13}$ for a coherent input state. Therefore the LG correlation function, which we note with a prime to indicate dephased input, takes the same form as Eq.~\eqref{eq:Kcoherent}:
\begin{equation}
    K'(\alpha) =  1 + 4e^{-\frac{3|\alpha|^2}{4}}\sinh(\frac{|\alpha|^2}{4}).
\end{equation}
This shows that the LGI is violated even if the input state is completely decohered, and underlines the fact that the LGI violation with a coherent state does not come the from quantum superposition involved in $\ket{\alpha}$ when represented in the Fock basis.

Interestingly, similar calculations with a different photon number probability distribution $q_n(\lambda)$ allow to compute $K'$ for a thermal state. Consider a right propagating thermal state input $\rho_1(\lambda) = \sum_{n=0}^{+\infty}q_n(\lambda)\dyad{0,n}$,
where $q_n(\lambda) = e^{-n\lambda}(1 - e^{-\lambda})$ and $\lambda = \hbar\omega / kT \in ]0,+\infty[$ defines the temperature through the photon energy $\hbar\omega$ and the Boltzmann constant $k$. As argued previously, $C_{13} = \ev{Q_3} = \Tr(\hat{Q}_3\rho_3(\lambda)) = - \sum_{n=0}^{+\infty} q_n = -1$, and $C_{23} = 0$.  Following a previous calculation, we also find $C_{12} = \sum_{n=1}^{+\infty} \frac{1}{2^{n-1}}q_n(\lambda) = 2(1 - e^{-\lambda})\left(\frac{1}{2e^{\lambda} - 1}\right)$. This yields
\begin{equation}
    K'(\lambda) = 1 + 2(1 - e^{-\lambda})\left(\frac{1}{2e^{\lambda} - 1}\right).
\end{equation}
The LG correlation function reaches its maximum $K_M = \frac{2}{(1 + \sqrt{2})^2} + 1 \approx 1.343$ at $\lambda_M = \ln(1+1/\sqrt{2})$, which shows LGI violations to be allowed, in theory, with thermal states.

\section{Intermediate dephasing}\label{sec:decoh}

Let us now examine the effect of decoherence after the first beam splitter. To do so, we choose to write down the intermediate state after the first beam splitter when there is no intermediate measurement as $\rho_2(\gamma) = (\mathbbm{1}\otimes \Delta_{\gamma})\rho_2$, where $\Delta_{\gamma}$ is a dephasing channel which simply introduces a damping factor $\gamma\in [0,1]$ on the off-diagonal terms of the right intermediate mode, in the Fock basis. Formally, $\Delta_{\gamma}$ is a meta-operator with operator-sum (or Kraus) representation $\{\sqrt{\gamma}\mathbbm{1}, \sqrt{1-\gamma}\dyad{n}{n}_{n\in\mathbb{N}}\}$ where $\gamma$ is the damping factor. In particular, $\Delta_1$ is the identity meta-operator and $\Delta_0$ completely decoheres a quantum state. 

For an input Fock state $\rho_1(n)$ the intermediate state with decoherence reads
\begin{multline}
    \rho_2(n, \gamma) = \frac{1}{2^n}\sum_{k,\ell = 0}^n\sqrt{\binom{n}{k}\binom{n}{\ell}}(-1)^{k+\ell}\\ \times(\gamma + (1-\gamma)\delta_{k,\ell})\dyad{k, n-k}{\ell, n-\ell},
\end{multline}
where $\delta_{k,\ell}$ is a Kronecker symbol. 

The $C_{12}$ correlator is unaffected by decoherence, as shown in Appendix~\ref{sec:c12derivation}. Decoherence does not affect $C_{23}$ either, as the trials where $Q_2 = +1$ is measured negatively correspond to completely dephased states (all photons are in the same mode).

\begin{figure}[h]
    \centering
    \includegraphics[scale=0.55]{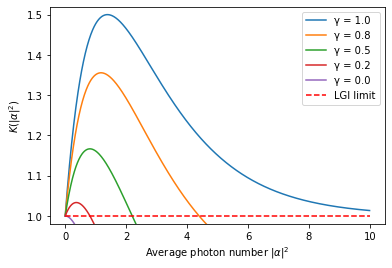}
    \caption{LG correlation function for coherent state or Poissonian Fock mixture input, with respect to the average photon number, for different values of the damping factor $\gamma$ of the intermediate dephasing.}
    \label{fig:Kfunctiondecohered}
\end{figure}

However, the decoherence affects $C_{13} = \ev{Q_3}$. With calculations similar to those shown in appendix~\ref{sec:proofs}, the relevant submatrix for the trace reads $\tilde{\rho}_3(n,\gamma) = (\gamma-1)\delta_{n,0}\dyad{0,0}{0,0} + \gamma\dyad{0,n}{0,n} + \frac{1}{4^n}\binom{2n}{n}(1-\gamma)\big(\dyad{n,0}{n,0} + \dyad{0,n}{0,n}\big)$. It follows that $\Tr(\Hat{Q}_3\tilde{\rho}_3(n,\gamma)) = (1-\gamma)\delta_{n,0} - \gamma -2\delta_{n,0} \frac{1}{4^n}\binom{2n}{n}(1-\gamma)$, and performing the weighted sum with the distribution given by Eq.~\eqref{eq:poissonianprob} gives $\ev{Q_3} = -e^{-|\alpha|^2} - \gamma(1 - e^{-|\alpha|^2})$. This results in a new LG correlation function of two variables, which plots for a few values of the damping factor $\gamma$ are shown in Fig.~\ref{fig:Kfunctiondecohered}, and which expression reads
\begin{equation}
    K'(\alpha,\gamma) = 4e^{-\frac{3|\alpha|^2}{4}}\sinh(\frac{|\alpha|^2}{4}) + (1-\gamma)e^{-|\alpha|^2} + \gamma.
\end{equation}
If the input state is a coherent state $\ket{\alpha}$, it turns out all the correlators are identical. For completeness, the derivation of those correlators can be found in Appendix~\ref{sec:bigproof}.\\

We observe that the LGIs are still violated even with strong decoherence. In fact, as long as the state after the first beam splitter is not completely decohered $(\gamma > 0)$, an LGI violation remains possible, and the only way an LGI violation is realized for all non-zero laser intensity is if there is strictly no loss of coherence $(\gamma = 1)$.

The explicit maximum with respect to the damping factor is reached when $|\alpha|^2 = 2\ln(1+\gamma)$. So we expect the maximal value to be reached at lower and lower laser intensities with increasing decoherence. The corresponding maximum reads
\begin{equation}
    K'(\alpha_{\text{max}},\gamma) = 1 + \frac{\gamma}{1+\gamma}.
\end{equation}
This final form is consistent with previous results, yielding no LGI violation for $\gamma = 0$ (complete decoherence) and a maximum of $1.5$ when $\gamma = 1$ (no decoherence).\\

\section{Conclusion}
 We have found that LGI violations can be achieved with classical states of light in an elementary Mach-Zehnder setup. We presented ideal negative measurements in the single photon case, and shown how to choose suitable observables values to use negative measurements for multiple photon numbers. In order to identify what was at the origin of the violation, we have considered coherent states with no phase reference, and examined the effects of decoherence after the first beam splitter.

Should the experiment be carried out successfully, this would be consistent with the fact that classical wave mechanics, being able to reproduce quantum random walk statistics~\cite{jeong2004simulation}, can simulate an LGI violation.
%Those violations, should they be observed experimentally, would show that macroscopic realistic descriptions of light, even classical, should be invalid, although they do become hard to invalidate in practice for high photon numbers, or for considerable intermediate phase noise.

\begin{acknowledgments}
This work was funded by the ESPRC Centre for Doctoral Training in Controlled Quantum Dynamics, and QuantERA ERA-NET Cofund in Quantum Technologies implemented within the European Union's Horizon 2020 Programme and KIAS visiting professorship. MSK acknowledges the Royal Society.
\end{acknowledgments}

%The LGIs and consequent questions about macroscopic realism will be of further interest in future works. Indeed, most efforts in nanotechnology and quantum information involve the preservation of quantum coherence to greater scales of system sizes, time spans and energy. Prospective experimental achievements may give rise to a true LGI loophole free test, on a genuinely macroscopic system. Future work on realism will also stem from questioning assumption \textbf{(I)} which might benefit from a conceptual shift regarding time and induction. In this sense, the analogy between Bell's inequalities and the LGIs could be at the center of future discussions, and indeed multiple extensions of the current quantum formalism~\cite{vedral2015quantum,oreshkov2012quantum,leifer2013towards} have already started to gain momentum.
\bibliography{ref}

\bibliographystyle{apsrev4-1}

\newpage
\onecolumngrid
\begin{appendix}

\section{Derivation of $C_{12}$ for decohered input}\label{sec:c12derivation}

We prove that $C_{12} = \sum_{n=1}^{+\infty} \frac{1}{2^{n-1}}p_n(\alpha)$ when the input state $\rho_1(\alpha) = \sum_{n=0}^{+\infty} p_n(\alpha)\dyad{n,0}{n,0}$ is the dephased coherent state in the left mode. The intermediate state $\rho_2(n,\gamma)$ is obtained by propagating a Fock state input $\rho_1(n) = \dyad{n,0}{n,0}$ through the beam splitter and dephasing in the right intermediate mode, so that
\begin{equation}
    \Tr(\hat{Q}_2\rho_2(n,\gamma)) = \frac{1}{2^n}\sum\limits_{k,\ell = 0}^n\sqrt{\binom{n}{k}\binom{n}{\ell}}(-1)^{k+\ell}(\gamma + (1-\gamma)\delta_{k,\ell})\mel{\ell,n-\ell}{\hat{Q}_2}{k, n-k}.
\end{equation}
From Eq.~\eqref{eq:q2def} one finds $\hat{Q}_2\ket{k,n-k} = (1-\delta_n)\left(\ket{0,n}\delta_k + \ket{n,0}\delta_{k,n}\right)$, hence the general matrix element reads $\mel{\ell,n-\ell}{\hat{Q}_2}{k, n-k} = (1-\delta_n)(\delta_{\ell}\delta_k + \delta_{\ell,n}\delta_{k,n})$. From this, one can deduce $\Tr(\hat{Q}_2\rho_2(n,\gamma)) = \frac{1}{2^{n-1}}(1-\delta_n)$, and the announced result follows immediately. Note that the dephasing parameter $\gamma$ does not affect this correlator.

\section{Derivation of projected output density matrix}\label{sec:proofs}

We prove that  $\forall n\in\mathbb{N}, \ \tilde{\rho}_3(n) = \dyad{0,n}{0,n}$.

The intermediate state is 
\begin{equation}
    \rho_2(n) = \frac{1}{2^nn!}\sum\limits_{k,\ell=0}^n \binom{n}{k}\binom{n}{l}(-1)^{k+\ell}\hat{a}^{\dagger\,k}_{\text{L}}\hat{a}^{\dagger\, n-k}_{\text{R}}\dyad{0,0}{0,0}\hat{a}_{\text{L}}^{\ell}\hat{a}_{\text{R}}^{n-\ell}.
\end{equation}
Applying the beam splitter transformation~\eqref{eq:BStransformations} yields
\begin{equation}
\begin{split}
    \rho_3(n) = \frac{1}{2^{2n}n!}\sum\limits_{\substack{k,\ell=0}}^n\binom{n}{k}\binom{n}{\ell}(-1)^{k+\ell}(\crea{a}_{\text{L}}-\crea{a}_{\text{R}})^k(\crea{a}_{\text{L}}+\crea{a}_{\text{R}})^{n-k}\dyad{0,0}{0,0}(\Hat{a}_{\text{L}}-\Hat{a}_{\text{R}})^{\ell}(\Hat{a}_{\text{L}}+\Hat{a}_{\text{R}})^{n-\ell},
\end{split}
\end{equation}
which can be expanded into
\begin{multline}
        \rho_3(n) =  \frac{1}{2^{2n}n!}\sum\limits_{k,\ell=0}^n\binom{n}{k}\binom{n}{\ell}(-1)^{k+\ell}\sum\limits_{i=0}^k\sum\limits_{j=0}^{\ell}\sum\limits_{p=0}^{n-k}\sum\limits_{q=0}^{n-\ell}\binom{k}{i}\binom{\ell}{j}\binom{n-k}{p}\binom{n-\ell}{q}\\
         \times (-1)^{k+\ell- i-j}\sqrt{(i+p)!(n-i-p)!(j+q)!(n-j-q)!}\dyad{i+p,n-i-p}{j+q,n-j-q}.
\end{multline}
Then in the evaluation of $\Tr(\hat{Q}_3\dyad{i+p,n-i-p}{j+q,n-j-q})$ the only non-vanishing terms satisfy $(p = n-i)\wedge (q = n-j)$ or $(q = -j)\wedge (p = -i)$. Note that both cases account for $n=0$. Hence, the relevant submatrix for the calculation of $\ev{\hat{Q}_3}$ is deduced to take the simple form
\begin{equation}
    \tilde{\rho}_3(n) = \frac{1}{2^{2n}}\left[\dyad{n,0}{n,0}\left(\sum\limits_{k,\ell = 0}^n \binom{n}{k}\binom{n}{\ell}(-1)^{k+\ell}\right) + \dyad{0,n}{0,n}\left(\sum\limits_{k,\ell = 0}^n \binom{n}{k}\binom{n}{\ell}\right)\right] - \dyad{0,0}{0,0}\delta_{n},
\end{equation}
where a vacuum contribution was subtracted to correct for the $n=0$ case. Since the first double-sum equals $\delta_n$ and the second equals $2^{2n}$, the announced result is obtained.

\section{Correlators for coherent state input with intermediate dephasing}\label{sec:bigproof}
We consider the input state $\rho_1 = \dyad{\alpha}{\alpha}$. Then the intermediate state with dephasing can be written using Eq.~\eqref{eq:poissonianprob} as
\begin{equation}\label{eq:intermediate}
    \rho_2(\alpha,\gamma) = \dyad{\frac{\alpha}{\sqrt{2}}}{\frac{\alpha}{\sqrt{2}}}\otimes\left[\gamma\dyad{\frac{-\alpha}{\sqrt{2}}}{\frac{-\alpha}{\sqrt{2}}} + (1-\gamma)\sum_{n=0}^{+\infty}p_n\left(\frac{\alpha}{\sqrt{2}}\right)\dyad{n}{n}\right].
\end{equation}
It should be noted that the summed over element $\dyad{n}$ designates a Fock state, while $\dyad{\alpha/\sqrt{2}}$ are coherent states. The first correlator $C_{12} = \ev{Q_1Q_2} = \ev{Q_2}$ is
\begin{equation}\begin{split}
    \Tr[\hat{Q}_2\rho_2(\alpha,\gamma)] =& \sum_{n=1}^{+\infty} \gamma\left(\braket{0}{\frac{\alpha}{\sqrt{2}}}\braket{n}{\frac{-\alpha}{\sqrt{2}}}\braket{\frac{\alpha}{\sqrt{2}}}{0}\braket{\frac{-\alpha}{\sqrt{2}}}{n} + \{n \longleftrightarrow 0\}\right) \\&+ (1-\gamma)\left(\bra{0,n}\left(\dyad{\frac{\alpha}{\sqrt{2}}}{\frac{\alpha}{\sqrt{2}}}\otimes \sum_{m=0}^{+\infty}p_m\left(\frac{\alpha}{\sqrt{2}}\right)\dyad{m}{m}\right)\ket{0,n} + \{n\longleftrightarrow 0\} \right),
    \end{split}
\end{equation}
which can be simplified to
\begin{equation}
    \Tr[\hat{Q}_2\rho_2(\alpha,\gamma)] = \sum_{n=1}^{+\infty} 2\gamma e^{-\frac{|\alpha|^2}{2}}p_n\left(\frac{\alpha}{\sqrt{2}}\right) + (1-\gamma)2e^{-\frac{|\alpha|^2}{2}}p_n\left(\frac{\alpha}{\sqrt{2}}\right).
\end{equation}
Hence we find $C_{12} = 2e^{-|\alpha|^2/2}(1 - e^{-|\alpha|^2/2})$ which, as announced, is the same as with the depolarized case. This is not surprising as phase noise does not change the photon number statistics and there has not been any interference at this stage.

Let us now proceed with the $C_{13} = \ev{Q_3}$ correlator. The output state can be written as
\begin{equation}\label{eq:fulltrace}
    \rho_3(\alpha,\gamma) = \gamma(\dyad{0}{0}\otimes\dyad{-\alpha}{-\alpha}) + (1-\gamma)\sum_{n=0}^{+\infty}p_n\left(\frac{\alpha}{\sqrt{2}}\right)\hat{B}\dyad{\frac{\alpha}{\sqrt{2}}}{\frac{\alpha}{\sqrt{2}}}\otimes\dyad{n}{n}\crea{B}.
\end{equation}
In the calculation of $\Tr[\hat{Q}_3\rho_3(\alpha,\gamma)]$ the first term is unproblematic (gives $-\gamma$ as all the flux is in the right output mode). Let us focus on the second term where there is interference between a coherent state and a Fock state. We write $\hat{B}\ket{\frac{\alpha}{\sqrt{2}}}\ket{n} = e^{-|\alpha|^2/4}\sum_{k=0}^{+\infty}\frac{\alpha^k}{\sqrt{2^kk!}}\hat{B}\ket{k,n}$, so that
\begin{equation}\label{eq:trace}
    \bra{\frac{\alpha}{\sqrt{2}}}\bra{n}\crea{B}\hat{Q}_3\hat{B}\ket{\frac{\alpha}{\sqrt{2}}}\ket{n} = e^{-|\alpha|^2/2}\sum_{k,\ell =0}^{+\infty}\frac{\alpha^{*\ell}\alpha^k}{\sqrt{2^{k+\ell}k!\ell!}}\mel{\ell,n}{\crea{B}\hat{Q}_3\hat{B}}{k,n}.
\end{equation}
Standard calculations using Eqs.~\eqref{eq:BStransformations} and~\eqref{eq:q3def} give
\begin{equation}\label{eq:generalmel}
    \forall (\ell,m,k,n)\in\mathbb{N}^4, \ \mel{\ell,m}{\crea{B}\hat{Q}_3\hat{B}}{k,n} = -\delta_{\ell}\delta_m\delta_k\delta_n + \frac{\sqrt{(k+n)!(\ell+m)!}}{\sqrt{2^{k+n}k!n!\ell!m!}}(\delta_{\ell+m,k+n} - (-1)^{k+\ell}\delta_{\ell+m,k+n}),
\end{equation}
from which one finds
\begin{equation}\begin{aligned}
    \bra{\frac{\alpha}{\sqrt{2}}}\bra{n}\crea{B}\hat{Q}_3\hat{B}\ket{\frac{\alpha}{\sqrt{2}}}\ket{n} &= -e^{-|\alpha|^2/2}\delta_n.
\end{aligned}
\end{equation}
Hence $C_{13} = -e^{-|\alpha|^2} - \gamma(1 - e^{-|\alpha|^2})$.

Finally, as previously argued in the main text, $C_{23} = 0$ by the fact that $Q_3$ is on average null for trials where $Q_2 = +1$.
\end{appendix}
\end{document}